\documentclass[runningheads,a4paper]{llncs}

\usepackage{amssymb}
\usepackage{graphicx}
\usepackage{mathtools}
\usepackage{tikz}
\usetikzlibrary{shapes.geometric, arrows}
\usepackage{flowchart}
\usepackage{caption}
\usepackage{subfig}
\usepackage[misc]{ifsym}

\begin{document}

\mainmatter  
\captionsetup[figure]{labelfont={bf},labelformat={default},labelsep=period,name={Figure.}}

\title{Discovering Key Nodes in a Temporal Social Network}


%
%
%
%

\author{Jinshuo LIU\inst{1} 
\and Chenghao MOU\inst{2}
\and Donghong JI\inst{1} (\Letter)}

\institute{ School of National Cybersecurity, Wuhan University, China
\and Computer Science School, Wuhan University, China}

%
%

\maketitle

\begin{abstract}
\textbf{[Background]}Discovering key nodes plays a significant role in Social Network Analysis(SNA). Effective and accurate mining of key nodes promotes more successful applications in fields like advertisement and recommendation. \textbf{[Methods]} With focus on the temporal and categorical property of users' actions - when did they re-tweet or reply a message, as well as their social intimacy measured by structural embeddings, we designed a more sensitive PageRank-like algorithm to accommodate  the growing and changing social network in the pursue of mining key nodes.  \textbf{[Results]} Compared with our baseline PageRank algorithm, key nodes selected by our ranking algorithm noticeably perform better in the SIR disease simulations with SNAP Higgs dataset. \textbf{[Conclusion]} These results contributed to a better understanding of disseminations of social events over the network.

\end{abstract}

\section{Introduction}

Imagine a situation where one of your close friend told you a hilarious joke few weeks ago and you have made several friends by breaking the same jest since then. Later you come across the same joke coincidently on the humor columns in the newspaper. Should this particular newspaper take credit for your already-bounded friendship? Probably no. Even if you keep uttering the joke after that encounter, the newspaper would be accredited only for your new friends; but the originator of the joke should take most of the credits, if any, as a major contributor to your friendship bounding.  With the same token, the arrival time(few weeks ago vs. someday later), forms(personal oral representation vs. text in newspaper) of the information and relational closeness(close friend vs. random newspaper) combined serve as a cornerstone in assessment of its value, and furthermore evaluation of its creator's influence(friend vs. newspaper).

In general, discovering key nodes within a network can be regarded as a mission to acquire the structural or functional influence ranking in SNA\cite{SNA}  and identification of such nodes has been widely researched. More than thirty algorithms are proposed from a wide spectrum of perspectives\cite{intro}, such as neighbor-based(e.g., k-shell decomposition\cite{kshell}, degree centrality\cite{Degree}, \cite{WDegree}), path-based(e.g., closeness centrality\cite{Closeness}, Katz centrality\cite{Katz}), eigenvector-based(e.g., LeaderRank\cite{LR}\cite{WLR}, PageRank\cite{PR}), and sometimes even combined one\cite{MultipleCriterion}. Evidently, PageRank, one of the eigenvector centrality analysis approach, has witnessed great success in applications in social networks. The initial idea of PageRank bases upon the assumption that the more links one web page receives, the higher it ranks. More recently, revised versions of PageRank such as LeaderRank and their weighted versions are proposed to adjust the traditional algorithm for the social network. 

However, most of those algorithms solely concern the relational social network - links referring to who follows whom. It is often the case that Internet tweets created by "nobody"s suddenly go viral and in certain circumstances, normal people have more direct and immediate access to local news than medias. So it is crucial to identify those key nodes in a dissemination social network - links containing the annotation of actions(e.g., re-tweeting or replying). 

To address this issue, we in this paper target at the voting process in the traditional PageRank algorithm within a relatively compact time interval(few days). Our research uses time-slots to slice one's votes for its inbound-links. The process allows us to generate a matrix for harnessing its shares and a vector for total votes. The second emphasis of our research is to distinguish different types of actions. Different actions - re-tweeting or replying - conveys the same information with different intensity: obviously, re-tweeting is more public and therefore attracts more tweet-mongers.  Those actions are then assigned with different weights to discriminate the rankings. Moreover, as is shown in the story, the intimacy between users also affects the distribution of one's votes. Based on the Node2Vec\cite{Node2Vec} algorithm, we are able to generate node embeddings instantly for structural similarity - a simple but efficient representation of players' intimacy for our analysis. Our experimental results show that even tough with minor defect in time consumption, our ranking algorithm is more representative and accurate than the original LeaderRank and PageRank. 

This paper is organized as follows. Section 2 discusses previous work that relates to our research.  Section 3 formally presents our approach and elaborates specifications and steps in the algorithm. Section 4 describes the data that we use, presents our experimental results, and shows simulations as our evaluation method. Finally, Section 5 summarizes our findings and conclusions.

\section{Related Works}
Much previous work on ranking nodes in social context has primarily focused on relational network and so far algorithms like PageRank\cite{PR} and LeaderRank\cite{LR},\cite{WLR} give a reasonable answer about the ladder of users' social influence. In typical scenarios, those algorithms are applied to a static social network and they tell us about the relative authoritative scores of each user, regardless of information flows that happen upon the network.

There are some existing works exploring the temporal information: Wan utilizes the decaying value of information over time for web page predication\cite{TimeRank2}. They categorize the time-and-activity relationship into four types in order to assess the information value from a broad time interval(few weeks to months); Fiala applies the temporal attribute to a citation network with especial attention on publication time and co-authorship and thus weigh the citations more discriminatingly\cite{CitationRank}. Though they have taken time into consideration, they did not touch on the actions and the corresponding effect of when and how users take their actions. The problem of  ranking user nodes within a propagation social network remains. 

\section{Method}

In this section, we introduce the proposed NodeRank algorithm for identifying key nodes in a temporal social network(dissemination graph). We first give the problem definition and notations. The we give an overview of our method. Finally, we reveal the details for each component in our approach as well as some calculation tricks to speedup the iteration.  

\subsection{Problem Definition and Notations}

A set of nodes $N_R = \{x_i| 0 \leq i \leq |N_R|\}$ and a set edges $E_R = \{e_j| 0 \leq j \leq |E_R|\}$ forms the relational network denoted as $G_R$ and the dissemination event is a subgraph of the relational network denoted as $G_D = \{n_i|0 \leq i \leq |N_D|\} \subseteq G_R$ where nodes are involved in propagating the information by various actions. $|N_R|$ and $|E_R|$ represent the number of nodes and edges in the relational network correspondingly, so do $|N_D|$ and $|E_D|$ in the dissemination network. Edges in the dissemination network $G_D$ is tagged with the time-stamp indicating exactly when the action takes place.  The task of discovering key nodes within the dissemination network is to give a ranking list of node influence within $G_D$ with or without extra information from relational network $G_R$. 

\subsection{An Overview of NodeRank}
In this section, we will briefly give an overview of our algorithm for identifying key nodes.

Our method could be  essentially regarded as an extension of the original PageRank algorithm. The tradition PageRank is an algorithm based on link popularity: the more links point to the page, the higher it ranks, as shown by the following equation.
$$PR(p_i) = \frac{1 - d}{N} + d\sum_{p_j \in G(p_i)}\frac{PR(p_j)}{L(p_j)}$$
Here, $p_i \in G_R$ is a web page in the collection we are going to rank. $G(p_i)$ is the collection of pages linking to $p_i$ and $L(p_j)$ is the number of out degree of page $p_j$. $d$ is the damping factor for handling non-strongly connected networks.

\begin{figure}[t]
\centering
\begin{tikzpicture}[font={\sf \small}]
\def \smbwd{1.5cm}
\thispagestyle{empty}
\tikzstyle{arrow} = [->,>=stealth]
\tikzstyle{line} = [-]

\node (start) at (0,0) [draw, terminal,minimum width=\smbwd, minimum height=0.5cm] {Start}; 
\node (getdata1) at (0,-1.5) [draw, predproc, align=left,minimum width=\smbwd,minimum height=1cm] {Temporal Network};
\node (getdata2) at (5,-1.5) [draw, predproc, align=left,minimum width=\smbwd,minimum height=1cm] {Network Embeddings};
\node (process1) at (2.5,-3.5) [draw, process, minimum width=\smbwd, minimum height=1cm] {Table and Vector Construction};
\node (decide) at (2.5,-5.5) [draw, decision, minimum width=\smbwd, minimum height=1cm] {Stop?};
\node (process2) at (2.5,-7.5) [draw, process, minimum width=\smbwd, minimum height=1cm] {Walk and Iteration};
\node (process3) at (5,-7.5) [draw, process, minimum width=\smbwd, minimum height=1cm] {Output};
\coordinate (point1) at (2.5,-8.75);
\coordinate (point2) at (2.5, -4.25);
\coordinate (point3) at (0,-7.5);
\node (end) at (5,-9.75) [draw, terminal,minimum width=\smbwd,minimum height=0.5cm] {End};

\draw[arrow] (start) -- (getdata1);
\draw[arrow] (getdata1) -- (getdata2);
\draw[arrow] (getdata1) |- (process1);
\draw[arrow] (getdata2) |- (process1);
\draw[arrow] (process1) -- (decide);
\draw[arrow] (decide) -| node[above]{Yes} (process3);
\draw[arrow] (decide) -- node[left]{No}(process2);
\draw[arrow] (process3) -- (end);
\draw[line] (process2) -- (point3);
\draw[arrow] (point3) |- (point2);

\end{tikzpicture}
\caption{The overall architecture of the algorithm.}
\label{Figure:1}
\end{figure}
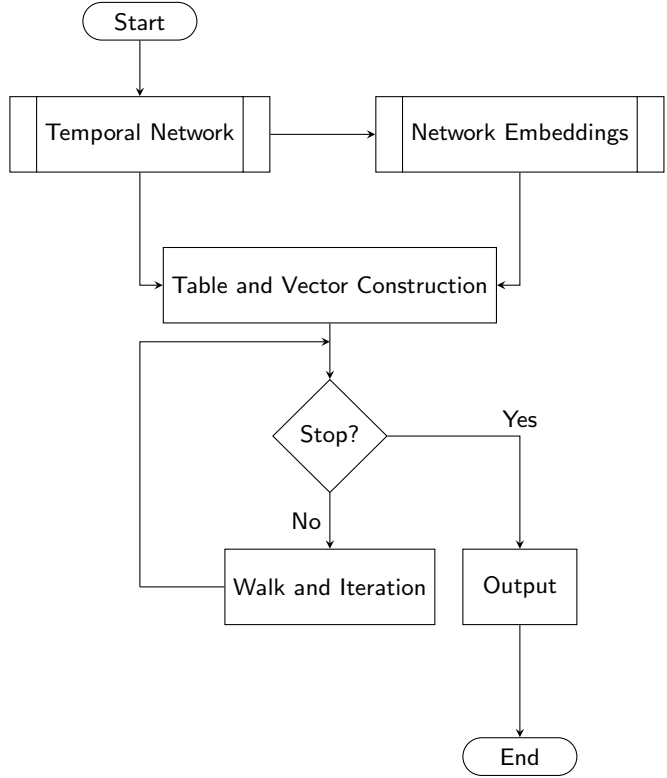

As is shown by the right hand side of the equation, each web page(e.g., $p_j$) evenly distributes its PageRank value to its informer $p_i$. In our revision of this formula, we follow several rules to capture useful information as distribution preferences for each node: (1) The earlier the information comes, the more votes(or values) the receiver will give to the informer; (2)The more commitment needed in an action, the more votes the receiver will give to the informer; (3) The closer the receiver and the informer are, the more votes the receiver will give to the informer.  In order to achieve this goal, we propose a base distribution table for each node with time in mind. Each table is sliced into time-based slots; each slot is filled with informers hitherto known to the node and broadcasters during this particular slot. What the node will do is to harness votes from those broadcasters and to potentially evenly distribute them to its informers - the oldest informers might take several round of votes. Upon the table, different actions and degree of intimacy will also change how much of the offered votes the informer receives. The overall architecture of the method is shown in Figure \ref{Figure:1}. 

Given a social network edge list containing all edges $E_R$ and a action list containing all featured edges $E_D$,  we first map edges to network graphs $G_R$ and $G_D$. In order to utilize the relational network, we will need to embed all node within $G_R$ first with a Node2Vec model. Then we initialize each node $n_i$ with a distribution table(matrix) $F_i \in \mathbb{R}^{L(n_i)\times T(n_i)}$, a collection vector $P_i \in \mathbb{R}^{T(n_i)}$, and an offering vector $O_i \in \mathbb{R}^{L(n_i)}$, where $T(n_i)$ is the number of time slots node $n_i$ has during this event. As indicated by the dimension superscript, distribution matrix $F_i$ is a coefficient matrix where each element $f^i_{pq}$ means the share of votes informer $n_p$ expects at time slot  $q$;this is where the action type factor and node embeddings come into play. As for the collection vector, each element of $P_i$ stores the total votes it collects at each time slot. Each element in $O_i$ stores the votes it will give to its corresponding informer.
As illustrated by Figure 1, after initialization, we will apply a random walk iteration until certain criteria are meet such as maximum iteration or minimum threshold for difference between iterations. During each iteration, a node will refresh its collection vector by querying its contributors/broadcasters and update its offering vector by applying $F$ to $P$. The final output of the algorithm will be an ordered list of summation of each node's collection vector. 

\subsection{Components}
In the following subsections, we will introduce all components of our algorithm. 
\subsubsection{Action Type Factor $\omega$}

Intuitively, different forms of participation should be attached with different weights considering a re-tweet is more time and energy consuming than a simple reply. Even though both actions are public to others in the social network, re-tweets are far more noticeable for potential tweet-mongers. Therefore, we formally categorize the actions into two types: re-tweeting and replying, each with unique weight $\omega_{rt}$ and $\omega_{rp}$, $\omega_{rt} + \omega_{rp} = 1$. Ideally, $\omega_{rt} \gg \omega_{rp}$. Each edge will be weighted according to its receiver's action, which will be denoted by $w_{i, j}$ - the action performed by node $n_j$ on the information given by node $n_i$.

\subsubsection{User Intimacy Factor $\mu$}
 
 Similar to the action type factor, we adopt one node embedding algorithm as our measurement of user intimacy. Presumably, similar users or user cliques share certain common interests that are reflected by a similar relation web; the Node2Vec algorithm "learns low-dimensional representations for nodes in a graph by optimizing a neighborhood preserving objective", which satisfies our need to preserve the similarity of the relation network. Each pair of nodes $i$ and $j$, the user intimacy factor is given by the cosine similarity between node vectors.
 $$\mu_{<i, j>} = \frac{V_i\cdot V_j}{||V_i||||V_j||}$$
 Where $V_i$ and $V_j$ are node embeddings for node $n_i$ and $n_j$.
 
\subsubsection{Distribution Table $F$ and Vectors}
As mentioned above, each edge in the dissemination network $G_D$ is both time- and type- sensitive. The information flows from informers to receivers and from receivers to broadcasters. Before we jump to the construction of the distribution table, we need to normalize those timestamps to evade the exhausting calculation since Internet timestamps are normally accurate to one second. The normalized time-stamp is expressed by the following equation.
$$T{(i, j|\beta)} = \frac{t_{i \to j}}{\beta}$$
Where $t_{i \to j}$ is the exact time-stamp for the action taken by $n_j$ for tweet originated from $n_i$ and $\beta$ is the normalization factor.

For each node $n_i$, it follows the following steps to initialize its settings:
\begin{enumerate}
 
\item Its offering vector is initialized with the value of $\frac{1 - d}{|N_D|\times L(n_i)}$.
\item Its time-slots $\tau(n_i) = \{t_m|0 \leq m \leq M_i\}$ are generated by orderly criss-crossing all timestamps of its normalized inbound and outbound links, $M_i$ indicating the number of time-slots for node $n_i$. 
\item Initialize the table $F$ with row index as its informers and column index as its ordered time-slots. The default value of the element remains zero. Initialize the collection vector $P$ with zeros and length set to the length of time-slots.
\item For each time slot $t$:
\item \begin{enumerate}
\item If there is no broadcaster at this time slot $t$, continue to the next one.
\item If there is no informers but there are any broadcasters at this time slot $t$, update the corresponding element in $P$ by collecting votes from those broadcasters' offering vectors.
\item Otherwise, there are both informers and broadcasters at this time slot $t$. Not only the collection vector will be update as in step 2, but each element $f^i_{pt}$ in $F$ will also be updated according to the formula $\frac{\omega_{<p, i>} \cdot \mu_{<p, i>} }{L^t(n_i)}$  if node $n_p$ is valid in this time slot.
\end{enumerate}
\end{enumerate}

More precisely, the initialization process can be defined as follows:
\begin{align}
  P_i &=        \begin{pmatrix} \sum_{n_j \in G^{t_0}(n_i)}O_j(n_i) & \sum_{n_j \in G^{t_0}(n_i)}O_j(n_i) & \cdots \sum_{n_j \in G^{t_{M_i}}(n_i)}O_j(n_i)& \end{pmatrix}         \\
  F_j & = 
  \begin{bmatrix}
    \frac{\omega_{<p_1, i>} \cdot \mu_{<p_1, i>} }{L^{t_0}(n_i)}       & \dots & \frac{\omega_{<p_1, i>} \cdot \mu_{<p_1, i>} }{L^{t_{M_i}}(n_i)} \\
    \hdotsfor{3} \\
    \frac{\omega_{<p_{L(n_i)}, i>} \cdot \mu_{<p_{L(n_i)}, i>} }{L^{t_0}(n_i)}        & \dots & \frac{\omega_{<p_{L(n_i)}, i>} \cdot \mu_{<p_{L(n_i)}, i>} }{L^{t_{M_i}}(n_i)} \\
\end{bmatrix}
  \\
  O_j         & = \begin{pmatrix} \frac{1 - d}{|N_D|\times L(n_i)} &  \frac{1 - d}{|N_D|\times L(n_i)} \cdots \frac{1 - d}{|N_D|\times L(n_i)}& \end{pmatrix}
\end{align}
Finally, we apply a random walk, visiting from node to node and updating its $P$ accordingly. At each visit, the offering vector $O$ will be updated by $F \cdot P$ as well.  After each iteration, we will check th following rules to measure the stability of the results. 

Compared with the traditional PageRank, an minimum threshold $\epsilon$ for results between iterations and maximum iteration number are both used in a short-circuit logic. Each iteration will produce a vector containing all node NodeRank values and we can calculate the difference between current and last iteration. Once the sum of the  difference is less than $\epsilon$ or the maximum number is reached, the algorithm outputs the current vector as the final result.

\section{Experiment}
\subsection{Settings and Dataset}
In our experiment for ranking node influence in a temporal network, we take two steps to analyze our proposed method. First, we implement the idea based on the available network dataset \cite{Higgs} from Standford Network Analysis Project, along with a direct comparison with the original PageRank algorithm. Next we put top-k nodes in both algorithms into modified SIR simulation trials to observe the average performance for each outcome. 

We used the largest strongly connected components in trials for each subgraph with only one action type or mixed action types after eliminating all redundant links-recording the earliest connection: Re-tweet network includes 984 nodes and 3850 edges in its largest strongly connected component, where there are 322 nodes and 702 edges in the reply network and 1801 nodes and 6601 edges in the mention network. More importantly, the mixed graph includes all three actions, 5548 nodes and 23378 edges in total.

In terms of evaluation, we used SIR simulations to test the influence of the nodes. During the evaluation, the whole dissemination network was sliced into time-series, simulating the dissemination process that the growing population was participating and initial seeds were selected from top-50 in both algorithms.

\subsection{Implementation Details}
Corresponding to our previous definition of action and user intimacy, our experimental settings for those factors are as follows: 1. When considered individually, each action is granted with $1.0$ weight and when mixed, $\omega_{rt(re-tweet)} = 1.0$, $\omega_{re(reply)} = 0.8$, $\omega_{mt(mention)} = 0.4$, and the default weight for unknown action is set to $0.6$. As for the node embeddings, dimension of each outcome is set to 64.The Normalization factor $\beta$ is set to $360$ minuets (roughly 6 hours) within the dissemination network.

Both ranking algorithms set the $\epsilon$ to $0.000001$ ans maximum iteration is set to 200. The damping factor remains the default value $0.85$ for all our trials.
\subsection{Results Comparison}
Formally, we will define our NodeRank model into three categories: NodeRank(NR for short, with all action and intimacy factors), NR-AI(without action and user intimacy factors), NR-A(without action factor) and NR-I(without user intimacy factor).
\setlength{\tabcolsep}{7mm}{
\begin{table}[htb]
\centering
\caption{Ranking Shifts Compared to PageRank}
\label{Table:1}
\begin{tabular}{ccccc}
\hline
\multicolumn{1}{c}{\textbf{Network}} 
& \multicolumn{1}{c}{\textbf{NR}}  
& \multicolumn{1}{c}{\textbf{NR-AI}} 
& \multicolumn{1}{c}{\textbf{NR-A}} 
& \multicolumn{1}{c}{\textbf{NR-I}} \\ \hline
Re-tweet                      & 192.12                                     & -  & -  & - \\
Reply                         & 63.83                                      & -  & -  & - \\
Mention                       & 364.26                                     & -  & -  & - \\
Mixed                         & 1111.09                                    & 1050.44 & 1056.57 & 1103.33 \\ \hline
\end{tabular}
\end{table}
}
Illustrated by table \ref{Table:1}, all trials have witnessed considerable shifts in node rankings comparing the baseline PageRank and our NodeRank. 

Furthermore, Figure \ref{Figure:2} shows a more vivid contrast between these two rankings. We ran the SIR simulation program 100 times, taking top-50 nodes in each method as initial seeds. 

Obviously, our selected nodes from NodeRank eventually infected more population than the traditional PageRank, approximately $\frac{3000}{600}=5$ times in the action-mixed network.

More importantly, Figure \ref{Figure:3} shows the comparison between our NodeRank(NR) and NodeRank-AI models. The slight higher infection rate in NodeRank indicates that both two factors have made joint efforts in depicting the network more discriminatingly; Figure \ref{Figure:3} and Figure \ref{Figure:4} manifest the performance between each factor: NR-I(Factor action) is better than NR-A(Factor user intimacy), and NR-A is better than NR-AI on average. More concisely, those figures verify our initial assumptions about those factors and each of them contributes positively to our final ranking results.

\begin{figure}[htb]
\centering
\includegraphics[width=.8\textwidth]{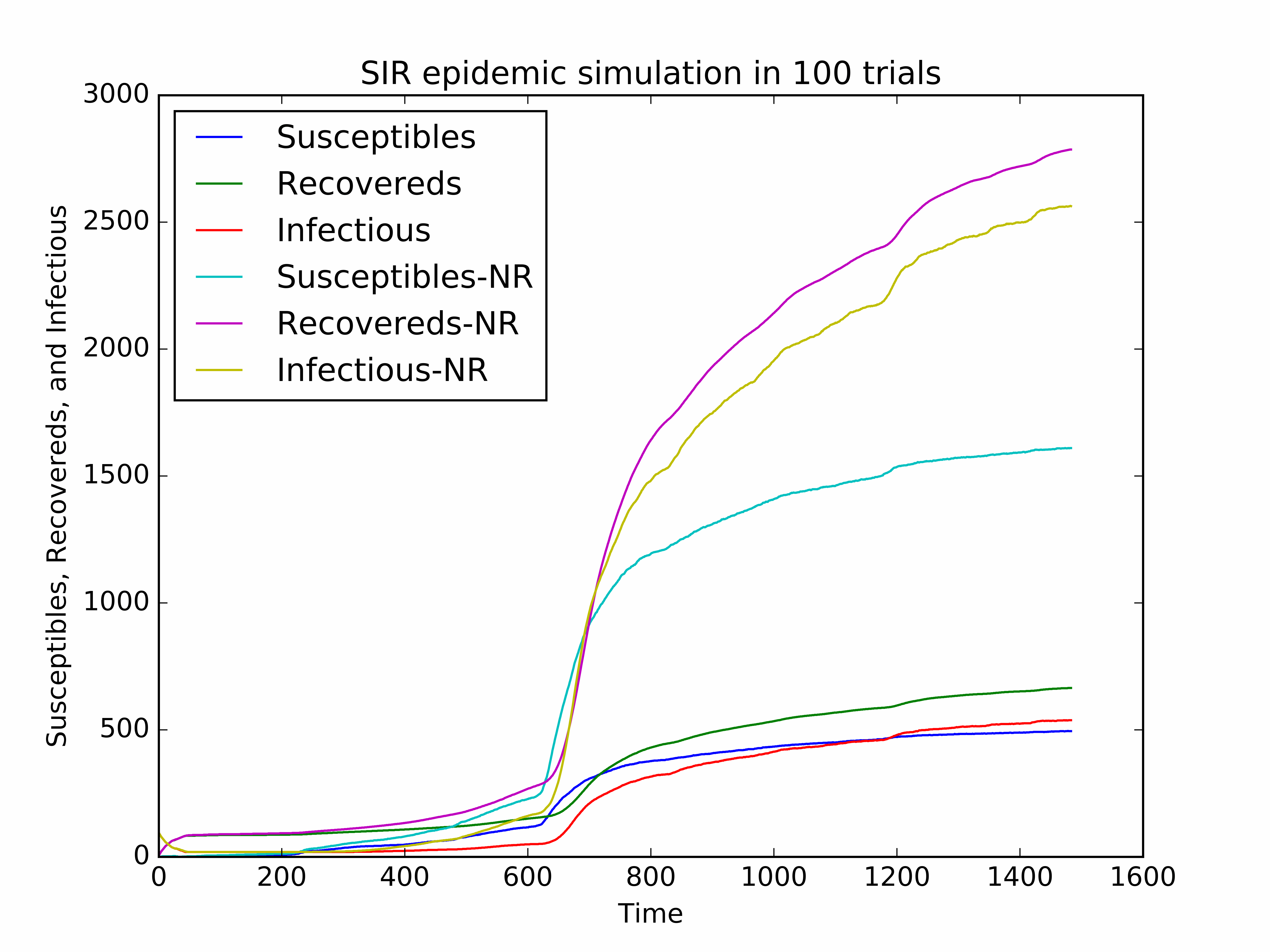}
\caption{Average SIR Growth Over 100 Trials in Mixed Graph}
\end{figure}
\label{Figure:2}

\begin{figure}[htb]
\centering
\begin{minipage}[t]{0.52\textwidth}
\centering
\includegraphics[width=7cm]{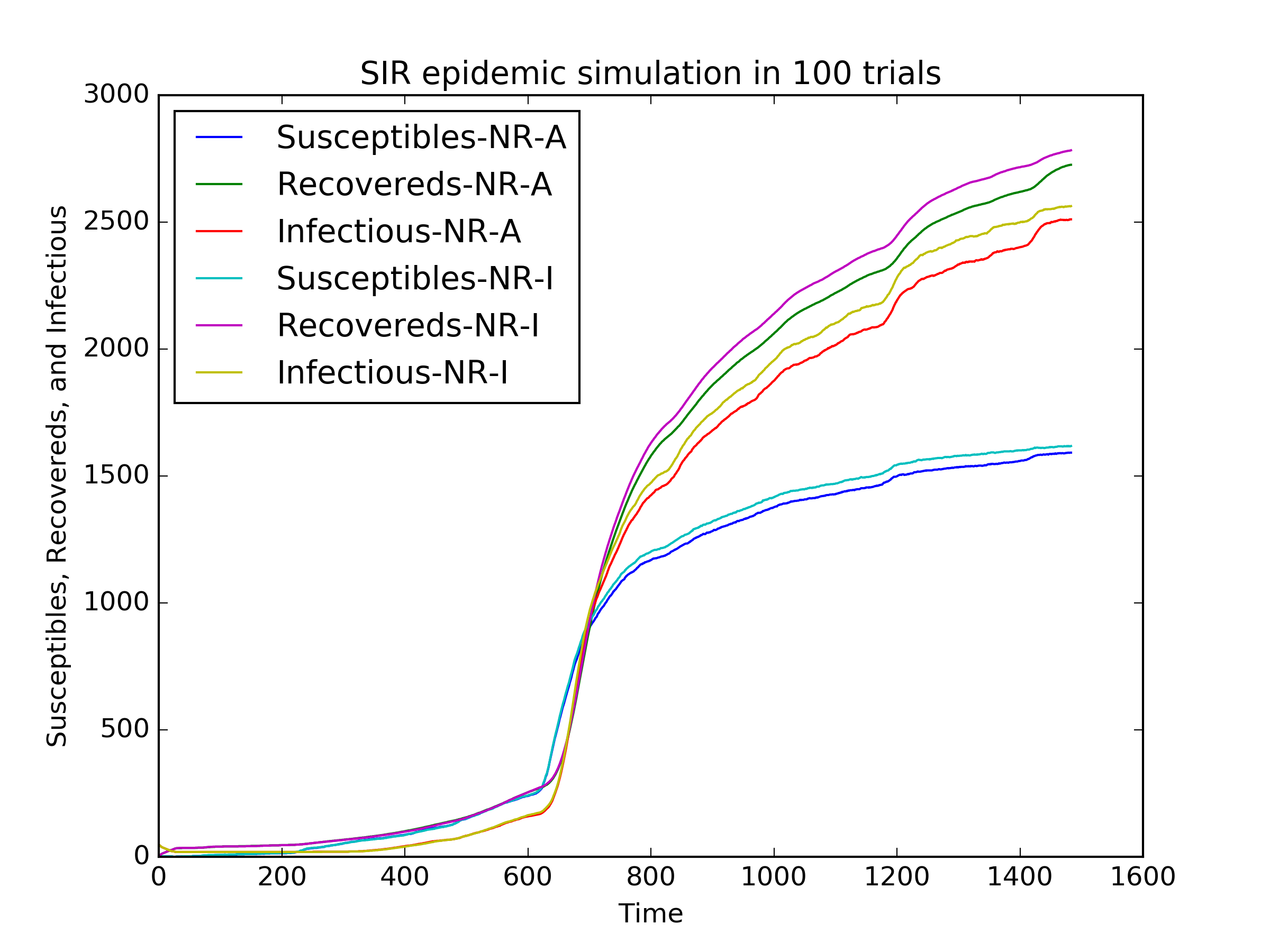}
\caption{Average SIR Growth Over 100 Trials in Mixed Graph Without Action VS. without Intimacy Factors}
\label{Figure:3}
\end{minipage}
\begin{minipage}[t]{0.45\textwidth}
\centering
\includegraphics[width=7cm]{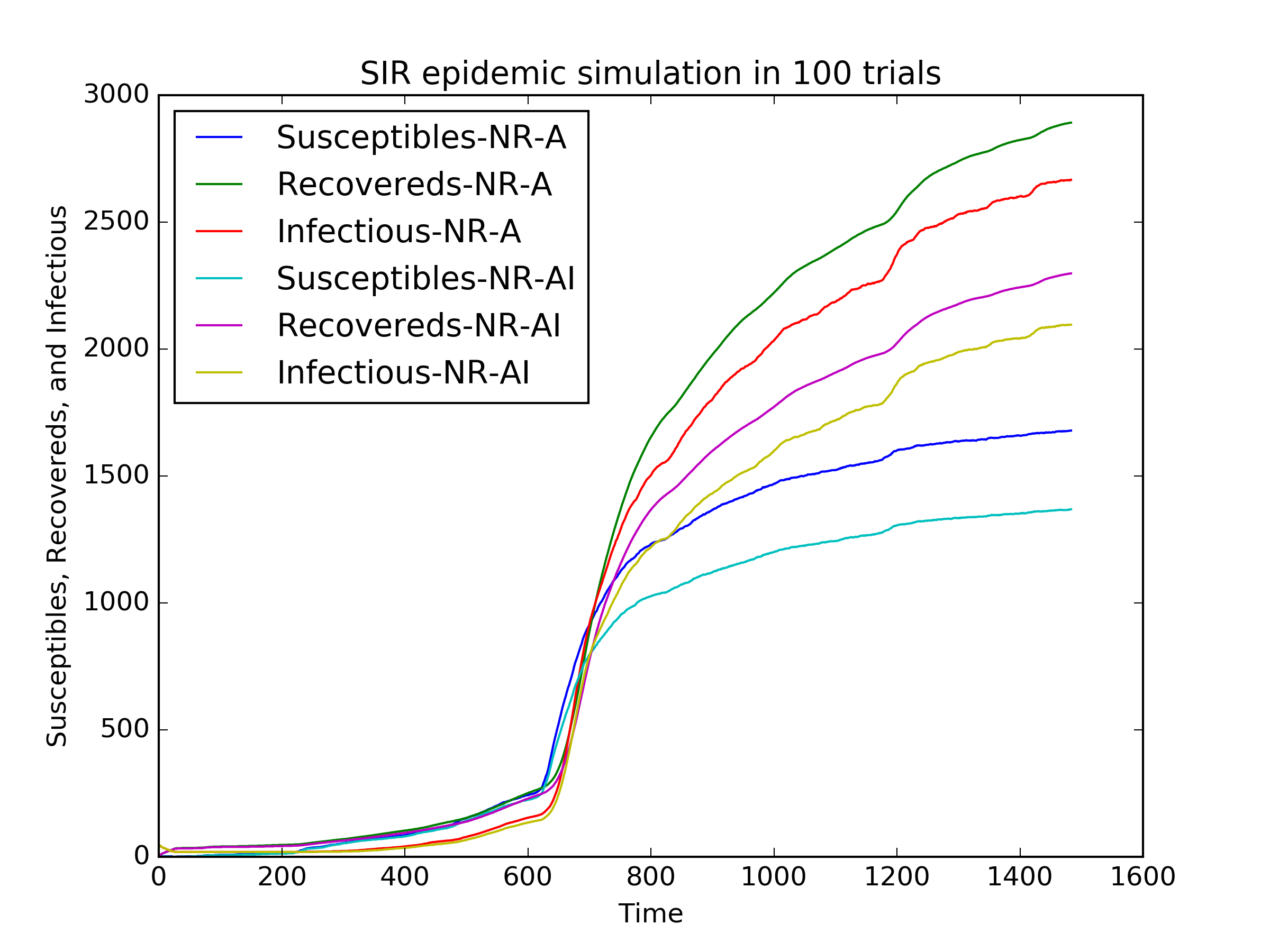}
\caption{Average SIR Growth Over 100 Trials in Mixed Graph Without Action VS. without Intimacy Factors}
\label{Figure:4}
\end{minipage}
\end{figure}

\begin{figure}[htb]
\centering
\includegraphics[width=.8\textwidth]{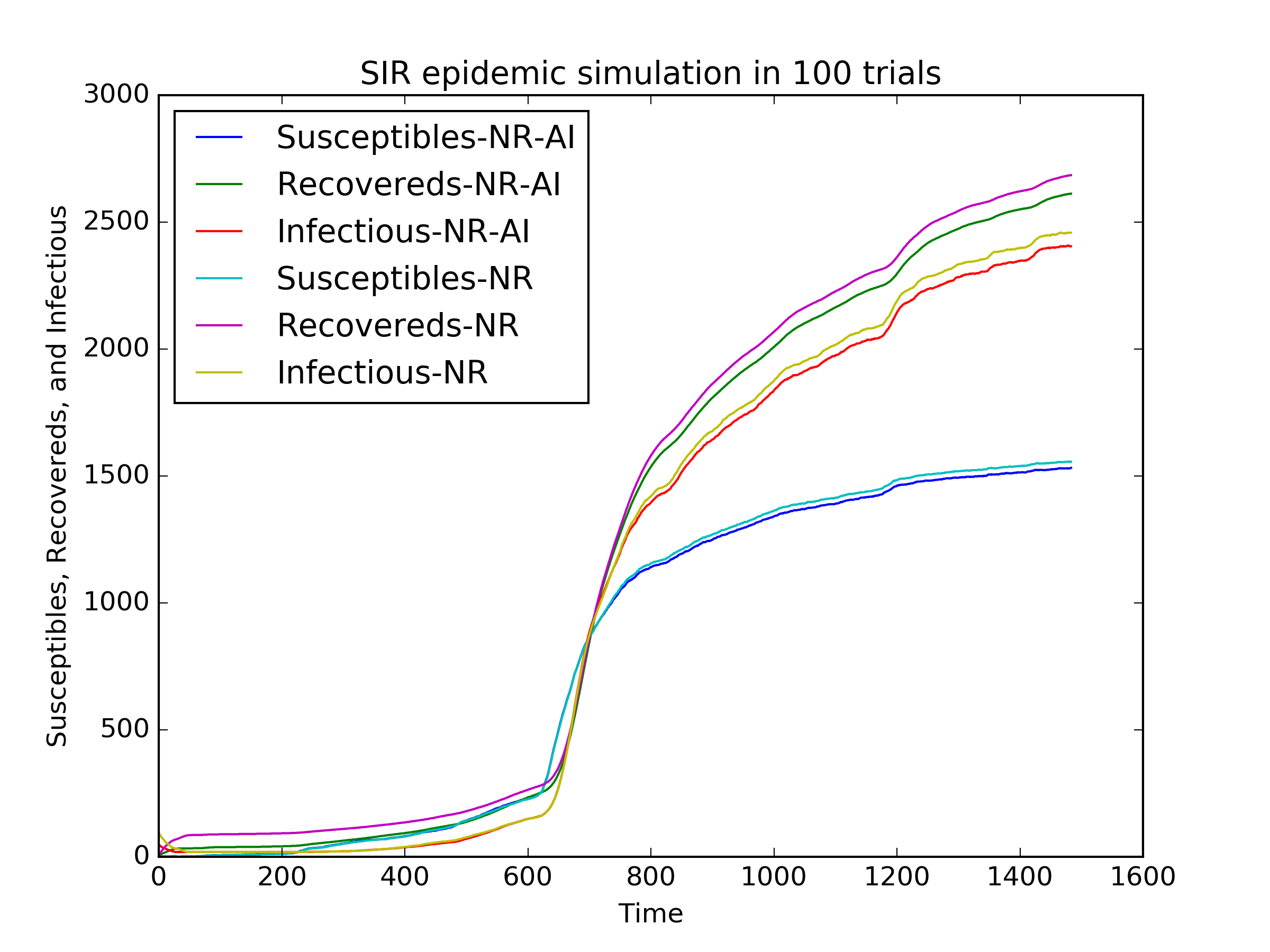}
\caption{Average SIR Growth Over 100 Trials in Mixed Graph Without Action and Intimacy Factors}
\end{figure}
\label{Figure:5}

\begin{figure}[htb]
\centering
\begin{minipage}[t]{0.52\textwidth}
\centering
\includegraphics[width=7cm]{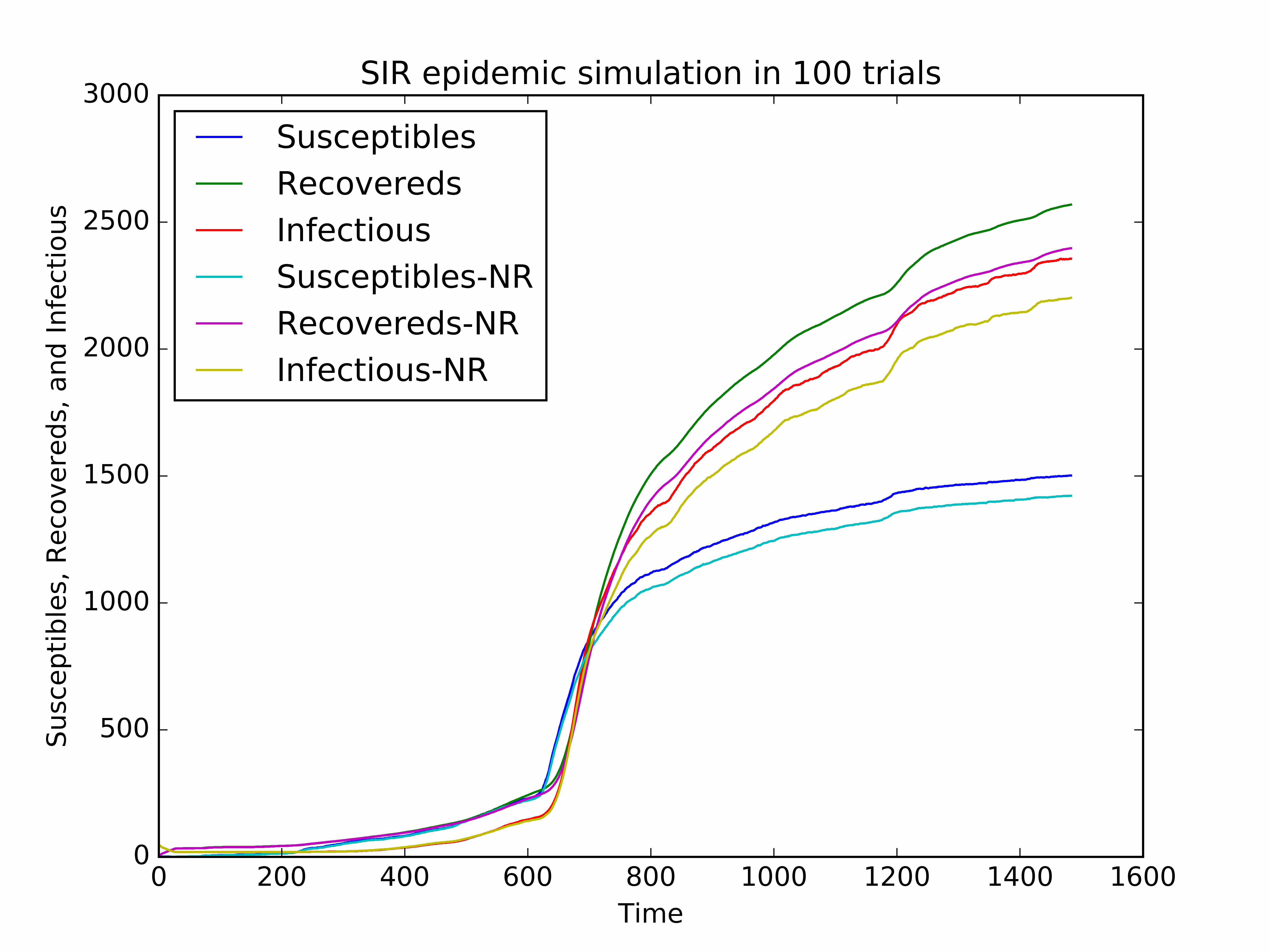}
\caption{Re-tweet Network}
\end{minipage}
\begin{minipage}[t]{0.45\textwidth}
\centering
\includegraphics[width=7cm]{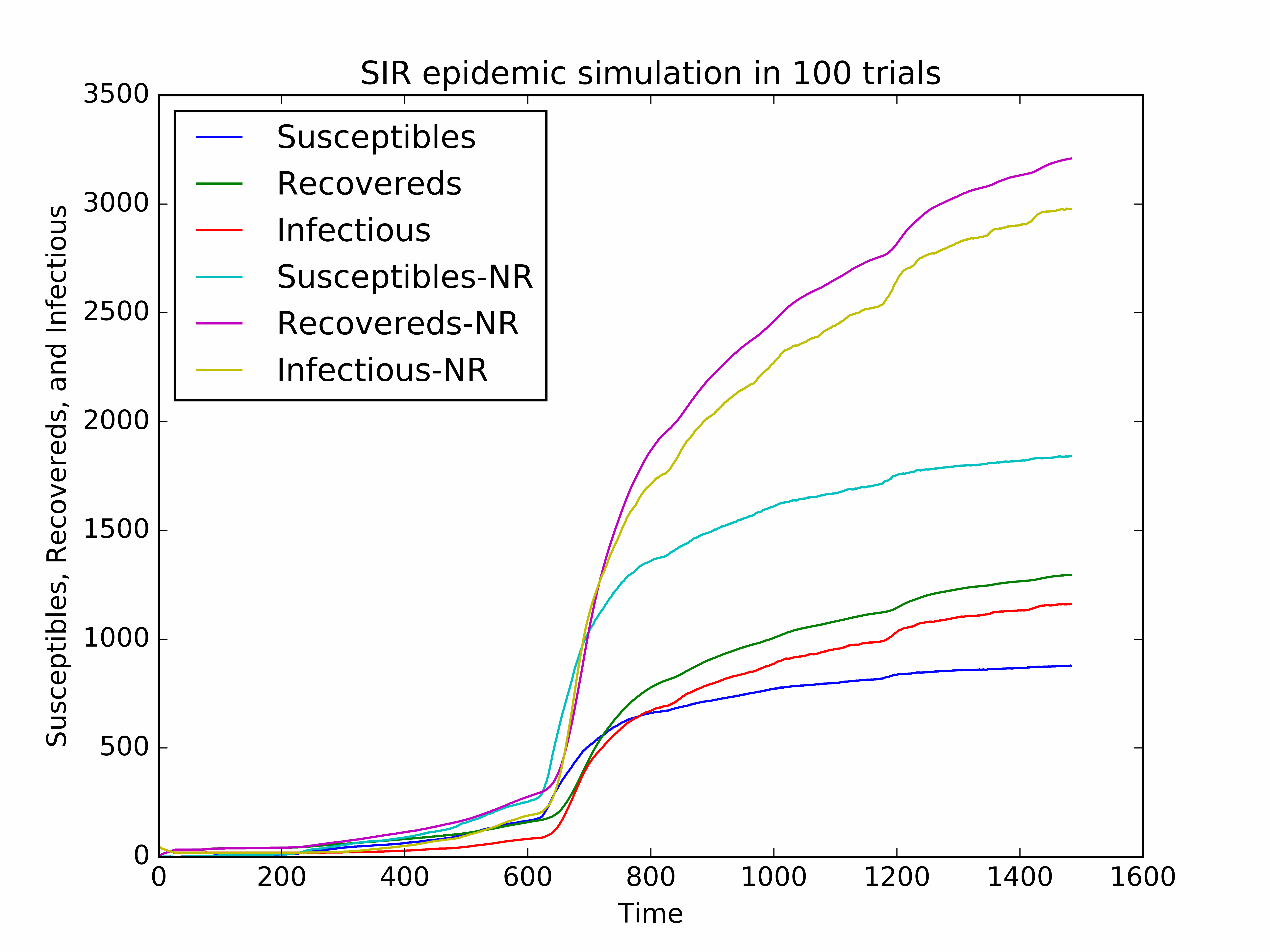}
\caption{Reply Network}
\end{minipage}
\begin{minipage}[t]{1\textwidth}
\centering
\includegraphics[width=7cm]{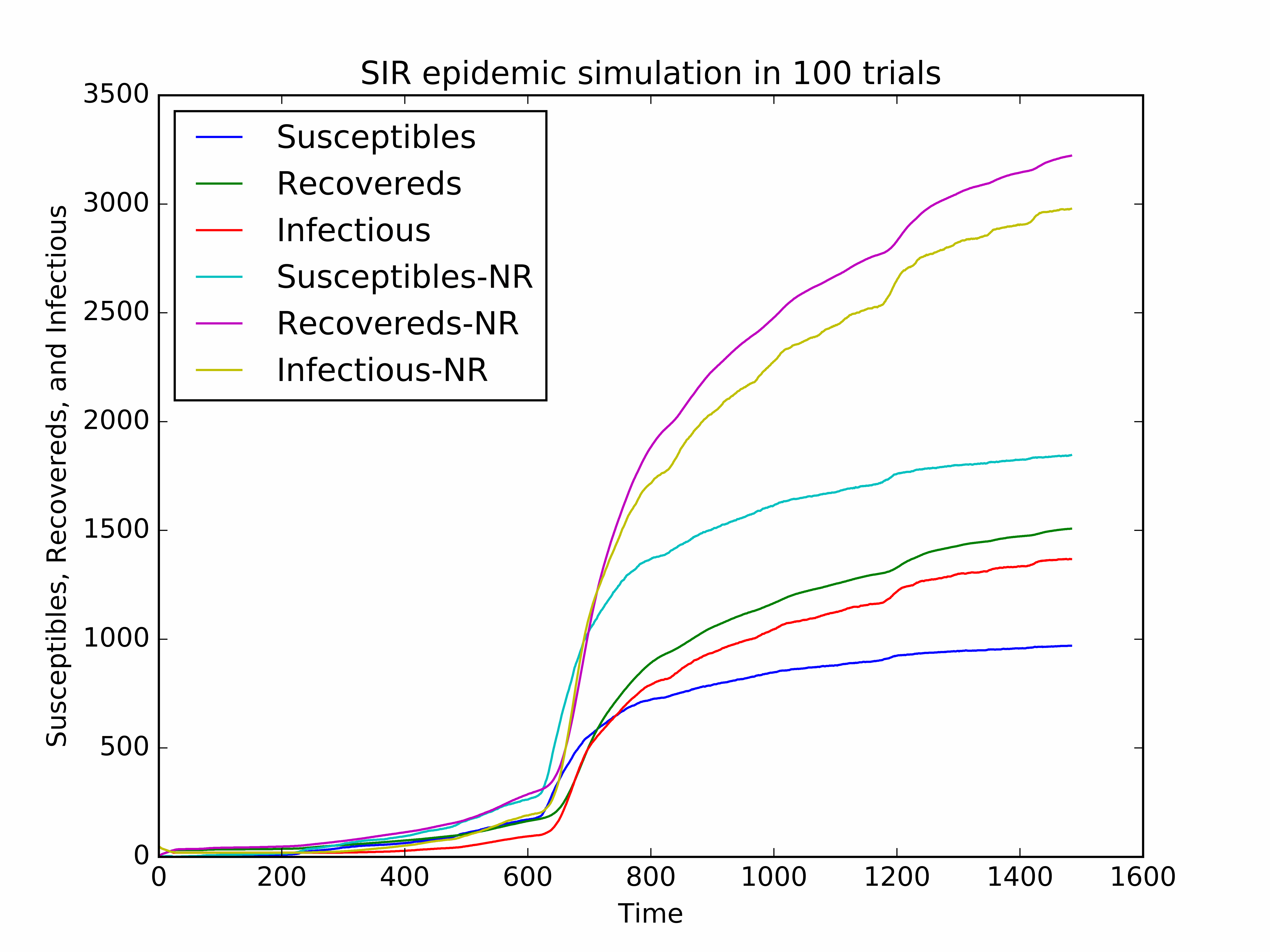}
\caption{Mention Network}
\end{minipage}
\caption{Average SIR Growth Over 100 Trials in One-action Networks}
\end{figure}

Though in the re-tweet network the results from NodeRank does not exceed those from PageRank with minor defeat, our approach outperformed the baseline in other networks.

\section{Conclusion}
Based on the temporal dissemination network, our NodeRank algorithm ranks the
influence of users with respect to their communication preferences including temporal, type information of each action as well as user intimacy. In
comparison with the conventional PageRank on the Higgs Data-set, our algorithms have provided a better, more accurate yet different results than
PageRank. However, it is still confronted with certain disadvantages in scalability compared to PageRank, and that is our future goal to remove such barriers.

\end{document}